\documentstyle[preprint,aps]{revtex}
\input{psfig}

\newcommand{\bra}[1]{\mbox{$\langle #1|$}}
\newcommand{\ket}[1]{\mbox{$|#1\rangle$}}

\begin{document}

\title
{{\bf   Meson retardation in deuteron  photodisintegration 
        above $\pi$-threshold}
  \footnote[2]
  {Supported by the Deutsche Forschungsgemeinschaft (SFB 201).}
}
\author
{M.\ Schwamb, H.\ Arenh\"ovel and  P.\ Wilhelm}
 \address{ Institut f\"ur Kernphysik,           
  Johannes Gutenberg-Universit\"at,  
  D-55099 Mainz, Germany }
\author{Th.\ Wilbois}
\address{Institut f\"ur Theoretische Physik, Universit\"at Hannover, 
 D-30167 Hannover, Germany}
\maketitle
\vspace{5mm}

\begin{abstract}
\noindent
Photodisintegration of the deuteron above $\pi$-threshold 
 is studied in a coupled
channel   approach including $N \Delta$- and  $\pi d$-channels  with
 pion retardation in potentials and exchange currents. 
 A much improved description of  total
and differential cross  sections in the energy region between $\pi$-threshold
  and 400-450 MeV is achieved.

\end{abstract}

\section{Introduction}

Despite its long history, photodisintegration of the deuteron is still
 a very  intensely studied
process. In the past decades, large efforts 
were made  in order to understand this simplest 
photonuclear reaction. 
At energies below $\pi$-threshold, where the theory is based on
 realistic $NN$-interaction models and corresponding exchange currents,  
 the agreement between theory and experiment is quite satisfactory
 \cite{{ArS91},{Lev95},{ScA91}}.
On the contrary, in the $\Delta$-resonance region where the 
 $\Delta$-excitation is dominant, the situation  is much less well 
 settled (for a review see \cite{ArS91}).
None of the models developed 
so far  such as the diagrammatic approach of Laget \cite{Lag78}, 
 the framework of nuclear isobar configurations in 
 the impulse approximation \cite{Sch95}, 
the  unitary three-body model \cite{TaO89} and the coupled channel 
 approach (CC) \cite{{LeA87},{WiA93}} is able 
to describe in a satisfactory manner the experimental data over  the whole 
 $\Delta$-resonance region \cite{{Leg95},{Cra96}}.
  In the impulse approximation \cite{Sch95} and also in the 
CC approach of \cite{LeA87},
 several parameters were adjusted
 in order to fit the total cross section whereas
 in the three-body model of \cite{TaO89} and in the coupled channel approach of
 \cite{WiA93} all  free parameters
 were fixed in advance  by fitting
 $NN$- and $\pi N$-scattering, and $\pi$-photoproduction on the nucleon. 
Consequently, no adjustable parameters remained for deuteron 
 photodisintegration. However, it turned out that both approaches 
 considerably underestimated the total cross 
section in the $\Delta$-region  by about 20-30\% \cite{{TaO89},{WiA93}}. 
Another failure was the wrong shape of the differential 
 cross section  which was flatter than the experimental one and which
 developed a more and more pronounced dip around $90^{\circ}$ with increasing 
 energy.
Also the  photon asymmetry was not well described,  especially at photon  
 energies above 300 MeV  \cite{{TaO89},{WiA93}}.
 
\begin{figure}
\centerline{\psfig{figure=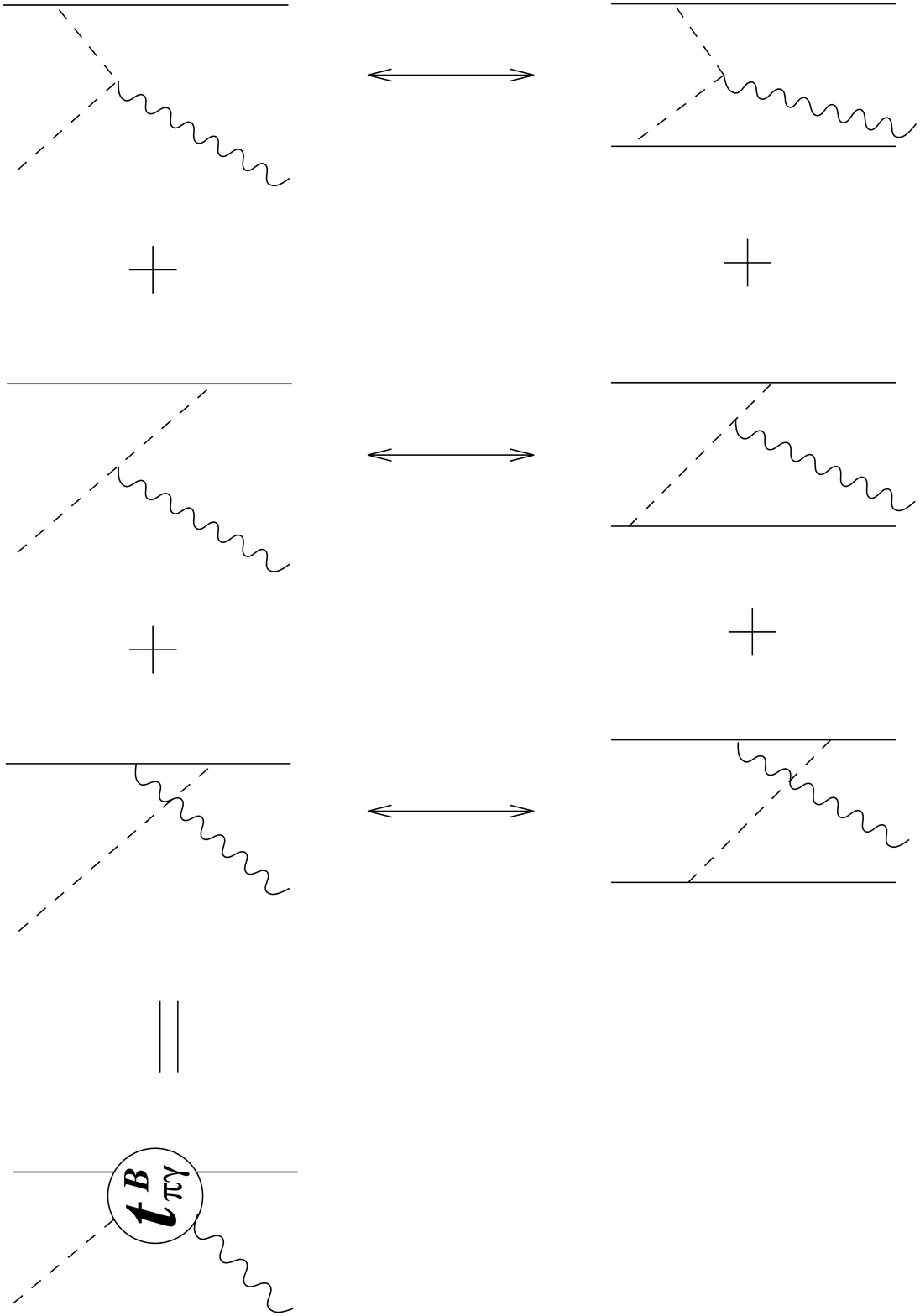,width=6.5cm,angle=270}}
\vspace{0.5cm}
\caption{The Born terms contributing to the 
$M_{1+}(\frac{3}{2})$-multipole amplitude of pion photoproduction 
(upper part) and their correspondence in the recoil and pion-MECs 
 in deuteron  photodisintegration (lower part).}
\label{vergleich}
\end{figure}

One of the crucial points in these calculations 
is the question of how to fix the 
$\gamma N \Delta$-coupling $G^{M1}_{\Delta N}(E_{\Delta})$ in the
 $M1 \,\,  N \Delta$-current 

\begin{equation}\label{ndcurrent}
\vec{\jmath}^{\,\,\, M1}_{\Delta N}(E_{\Delta}, \vec{k}) =
\frac{G^{M1}_{\Delta N}(E_{\Delta})}{2M} \,\, \tau_{\Delta N,0} \,\,
i\, \vec{ \sigma}_{\Delta N}
\times \vec{k} \, \, ,
\end{equation}
 where  $E_{\Delta}$ is the energy
 available for the internal excitation of the $\Delta$
 and $\vec{k}$  the momentum of the incoming photon.
While in the impulse approximation and also in \cite{LeA87} 
 the whole $P_{33}\,\, M1$-strength was 
 interpreted as
$\Delta$-excitation strength, the more sophisticated approaches 
 \cite{{TaO89},{WiA93}} have determined
 $G^{M1}_{\Delta N}$ by fitting the $M_{1+}(3/2)$-multipole of pion 
 photoproduction on 
 the nucleon including nonresonant Born terms, resulting in a smaller
  $G^{M1}_{\Delta N}$ and consequently a smaller photodisintegration cross
 section.  The Born terms
 contributing to the $(3,3)$-channel are the crossed $N$-pole and 
 $\pi$-pole graphs. 
 When embedded into the two-nucleon system, these Born terms become part
of the recoil and the $\pi$-meson current, respectively 
 (Fig.\ \ref{vergleich}).
In static calculations, however, the recoil current
  is not present due to its
 cancellation against the wave function renormalization current \cite{GaH76}. 
 A similar, but less  serious problem  arises in the treatment of
 the pion pole diagrams compared to the meson current of static MEC.
It had already been conjectured  in \cite{WiA93} that this inconsistent
 treatment
 of  pion exchange may lead to the observed underestimation of the total cross 
 section in their coupled channel approach, because by  incorporating 
 the Born terms effectively 
 into an increased   $M1\,\, \Delta$-excitation 
 strength a satisfactory agreement with the data could be achieved.

In the present  paper, we have  overcome
these shortcomings by  including for the first time in a coupled channel
 approach  complete retardation in the $\pi$-exchange  
 contributions to 
 potentials and MECs. This retardation is not approximated by keeping only the
 leading order contribution of a  $p/M$-expansion in the $\pi NN$-propagators
as was done for example in \cite{{ScA89},{GoA92}}. We furthermore consider the
 intermediate $\pi d$-channel in order to ensure all unitarity constraints
 up to the  $2 \pi$-threshold. 

\section{Theoretical framework}

In our framework, the Hilbert space is subdivided into three orthogonal
spaces ${\cal H} = {\cal H}_{N} \oplus {\cal H}_{\Delta} \oplus 
 {\cal H}_{Q}$ \cite{PoS87}.
${\cal H}_N$ consists of two nucleons only, in ${\cal H}_{\Delta}$ one of the
nucleons is replaced by a $\Delta$(1232)-isobar, 
and ${\cal H}_Q$ contains at least
one meson besides  two baryons. ${\cal H}_Q$ is then eliminated by introducing 
 effective operators in ${\cal H}_{N} \oplus {\cal H}_{\Delta}$
 using the projection operator technique.

For  photodisintegration  we need the outgoing 
$NN$-scattering wave with total energy $E$ which can be  expressed as
\begin{eqnarray}\label{NN-Gleichung}
\ket{NN^{(-)}} &= &
\left[ 1+ G_0^N(z) T_{NN}(z) + G_0^{\Delta}(z)  T_{\Delta N}(z)
\right] \ket{NN}\,\,, \quad z= E - i \epsilon \,\,. 
\end{eqnarray}
The $T$-matrix obeys a coupled
integral equation of Lippmann-Schwinger type
\begin{eqnarray}
T_{NN}(z) &=& \tilde{V}_{NN}(z) + 
\tilde{V}_{NN}(z) G_0^N(z)  T_{NN}(z)
+ \tilde{V}_{N \Delta}(z) G_0^{\Delta}(z)  T_{\Delta N}(z) \label{T-Matrix1}
 \,\, ,\\
 T_{\Delta N}(z) &=& \tilde{V}_{\Delta N}(z) + 
\tilde{V}_{\Delta N}(z)  G_0^N(z)  
 T_{NN}(z) + \tilde{V}_{\Delta \Delta}(z) G_0^{\Delta}(z) 
 T_{\Delta N}(z)\,\, . \label{T-Matrix2}
\end{eqnarray} 
Here, $G_0^N(z)$ denotes the free $NN$-propagator and
 $ G_0^{\Delta}(z)$
 the dressed  $N \Delta$-propagator, which depends on the
 bare $\Delta$-mass $M^{0}_{\Delta}$ and  the complex, energy dependent 
  $\Delta$-self energy
$\Sigma_{\Delta}$  taken from the dynamical model
 of Tanabe and Ohta (model A in \cite{TaO85}).

\begin{figure}
\centerline{\psfig{figure=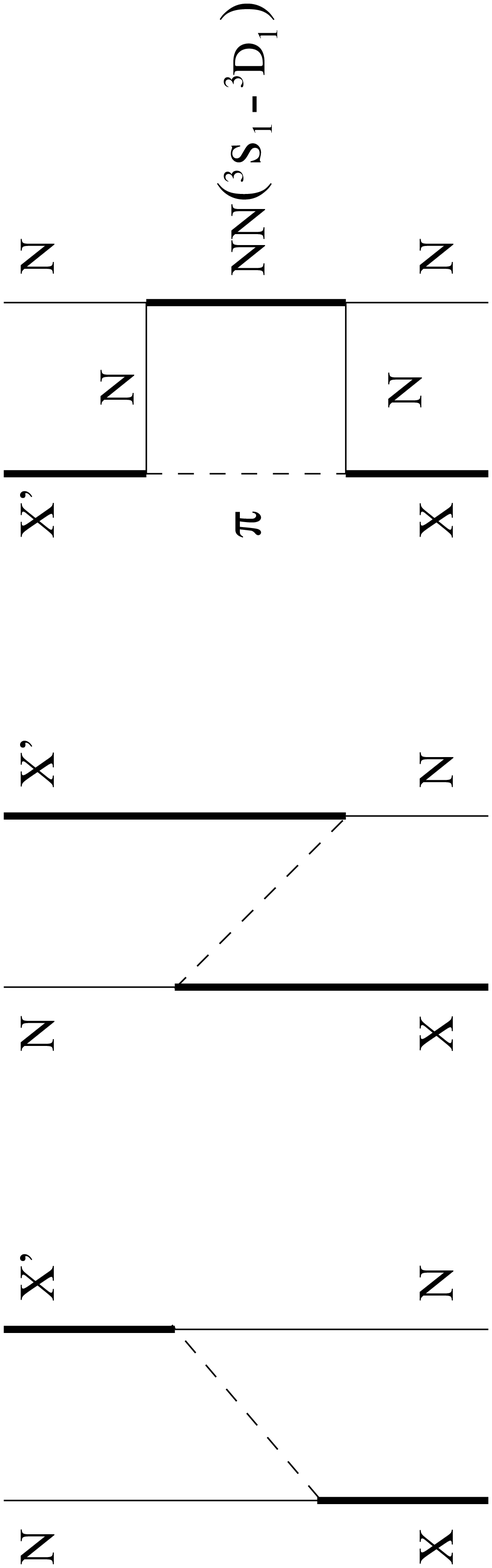,width=8cm,angle=270}}
\vspace{0.5cm}
\caption{Graphical representation of the hadronic interaction of Eq.\ 
 (\ref{VNN})  with  $X, X^{\prime} \in \left\{N, \Delta\right\}$.}
\label{hadint}
\end{figure}

The various effective driving potentials 
$\tilde{V}_{X X^{\prime}}$ with 
$X, X^{\prime} \in \left\{ N, \Delta \right\} $ in Eqs.\
 (\ref{T-Matrix1}) and   (\ref{T-Matrix2}) consist of two terms
(Fig.\ \ref{hadint})
\begin{equation}\label{VNN}
\tilde{V}_{X X^{\prime}}(z)=V_{X X^{\prime}}(z) + V_{X X^{\prime}}^{\pi d}(z)
 \,\, ,
\end{equation}
incorporating  respectively the usual boson-exchange mechanism and the
 formation of an interacting $NN$-pair with deuteron quantum numbers and 
 a pion as a spectator (denoted for simplicity by
 $\pi d$).
 In static applications, we have chosen  for $V_{NN}(z)$
 the  realistic,  
 energy independent Bonn-OBEPR model \cite{MaH87a}, whereas
 for the inclusion of full $\pi$-retardation  we have  used an improved 
 version of  the energy dependent Bonn-OBEPT model 
 developed by Elster et al.\,
  \cite{MaH88}
which contains besides retarded operators also self energy diagrams
calculated within a simple Lee model \cite{Lee54}.

 For $V_{N \Delta}$ and
$V_{\Delta N} = V_{N \Delta}^{\dag}$, we have constructed 
  corresponding static and retarded $\pi$- and $\rho$-exchange potentials, 
 taking for the $\pi NN$- and $\pi N \Delta$-form factors 
 the ones of the full
Bonn potential \cite{MaH87a}.
 The $\rho NN$- and $\rho N \Delta$-regulator
masses  are fixed  by fitting the $^1D_2$ 
$NN$-partial wave which is of crucial importance for deuteron 
 photodisintegration because of its strong coupling to the 
 dominant $^5S_2(N \Delta)$ partial wave. The diagonal interaction 
 $V_{\Delta \Delta}$ in  ${\cal H}_{\Delta}$
consists of the 'forward' and 'backward' going pion diagrams as depicted in
 Fig.\ \ref{hadint}. For its  $\pi N \Delta$-vertex, the
 form factor of \cite{TaO85} is used.

We now turn to the discussion of $V_{X X^{\prime}}^{\pi d}$,
 which  was studied in \cite{TaO89} but not included in \cite{WiA93}. 
From the corresponding diagram of Fig.\ \ref{hadint} 
 it  is obvious that $V_{X X^{\prime}}^{\pi d}$ can  be written as
\begin{equation}
V_{X X^{\prime}}^{\pi d}(z) = V_{X^{\prime} Q} G_0^{\pi N N, stat} 
t_{QQ}^{\pi d}(z) G_0^{\pi N N, stat} V_{Q X^{\prime}}\,\, ,
\end{equation}
where $V_{X^{\prime} Q}$ is the nonrelativistic $NN \pi$- or 
 $N\Delta \pi$-vertex, respectively.
 The  free $\pi NN$-propagator $G_0^{\pi N N, stat}$ 
 is taken in the static limit and  $t_{QQ}^{\pi d}(z)$ denotes 
the $NN$-scattering amplitude
in $Q$-space describing
an intermediate off-shell $NN$-state with the quantum numbers of the deuteron 
 and   a pion as spectator.  The amplitude $t_{QQ}^{\pi d}(z)$ is obtained 
  from the Lippmann-Schwinger equation
\begin{equation}
t_{QQ}^{\pi d}(z) =v^{\pi d} +  v^{\pi d} G_{0}^{\pi d}(z) 
 t_{QQ}^{\pi d}(z) \,\, ,
\end{equation}
where $G_0^{\pi d}$ denotes the free $\pi d$-propagator.
 It can be  solved easily by assuming a separable form for the driving
 term $v^{\pi d}$
\begin{equation}\label{vqq}
v^{\pi d} = \frac{V_{NN}(E^{\pi d}) \ket{d \pi; \vec{q}\,\,}
\bra{d \pi; \vec{q}\,\,}V_{NN}(E^{\pi d}) }{
\bra{d \pi; \vec{q}\,\,}V_{NN}(E^{\pi d}) \ket{d \pi; \vec{q}\,\,}} 
 \qquad \mbox{with} 
 \qquad E^{\pi d}=M_{d}+\frac{q^2}{4 M}
+ \sqrt{m_{\pi}^2+q^2} \,\, ,
\end{equation}
where $\ket{d \pi; \vec{q}\,\,}$ denotes a free $\pi d$-state with relative
momentum $\vec{q}$.
This ansatz for $v^{\pi d}$ satisfies 
for any realistic $NN$-potential $V_{NN}$
the Schr\"odinger equation
for the $\pi d$-state
\begin{equation} 
G_0^{\pi d}(E^{\pi d})v^{\pi d}\ket{d \pi; \vec{q}\,\,}=
\ket{d \pi; \vec{q}\,\,} \,\, .
\end{equation}

 Since the parameters of the realistic  
 $V_{NN}$ are  fitted to deuteron properties 
 and $NN$-scattering phases
 without explicit coupling to the $N \Delta$- and $\pi d$-channels, 
we have to apply a 
box-renormalization \cite{PoS87}
\begin{eqnarray}
 V_{NN}(z) &\rightarrow&  
 V_{NN}(z) - \tilde{V}_{N \Delta}(2M)  G_0^{\Delta}(2M)
\tilde{V}_{\Delta N}(2M) - V_{NN}^{\pi d}(2M) 
\end{eqnarray}
 in order to ensure an approximate phase equivalence in the 
 presence of these additional  channels.

 Now we turn to the e.m.\ part of our model. Above $\pi$-threshold,
 the  $\Delta$-excitation is the most important photoabsorption mechanism,
which is described by the current operator in Eq.\ (\ref{ndcurrent})
 neglecting small E2 contributions.
As mentioned above, $G^{M1}_{\Delta N}(E_{\Delta})$ 
is an  energy  dependent and complex coupling
which is fitted to the experimental
$M_{1+}(\frac{3}{2})$-multipole  of pion photoproduction on the
nucleon.  
 It contains besides  the bare $\gamma N \Delta$-coupling
 contributions from nonresonant pion  rescattering.
 The full pion photoproduction amplitude $t_{\pi \gamma}(E_{\Delta})$ 
 in the $(3,3)$-channel can be written as
\begin{equation}\label{deltat}
t_{\pi \gamma}(E_{\Delta}) = t^{B}_{\pi \gamma}(E_{\Delta}) -
 \frac{v^{\dag}_{\Delta} \vec{\epsilon} 
\cdot \vec{\jmath}^{\,\,\, M1}_{\Delta N}(E_{\Delta},\vec{k})}{
      E_{\Delta}-M^{0}_{\Delta} -\Sigma_{\Delta}(E_{\Delta})} \,\, ,
\end{equation}
where $t^{B}_{\pi \gamma}(E_{\Delta})$ is the nonresonant Born amplitude.
While in \cite{WiA93} 
an effective 
$\gamma  N \Delta$-coupling  $G^{M1}_{\Delta N}(E_{\Delta}) $
 and the model of   \cite{PoS87} for 
 the bare $\Delta$-mass  $M^{0}_{\Delta}$, the $\Delta$-self energy
 $\Sigma_{\Delta}$ and the $\Delta \pi N$-vertex
 $v^{\dag}_{\Delta}$ has been used,
 we follow here the work of Tanabe and Ohta (model A in \cite{TaO85}) 
 in which an explicit evaluation of the half-off-shell Born amplitude 
 in the $M_{1+}(3/2)$-multipole has been performed.

Furthermore, concomitant with the construction of the effective interactions,
 we have derived the corresponding effective two-body charge and current 
 density operators which fulfil current conservation  (for details, see
 \cite{Sch97}). These $\pi$-retarded MECs contain 
 besides the usual vertex-, meson- and 
 contact-MECs the recoil current and  charge densities 
  and a couple of additional  two-body  operators
 which  vanish identically in  the static limit.
Whereas the  latter ones 
 yield only  very small contributions which can be safely 
 neglected, the recoil contributions  turn out to 
 be quite important (see discussion below).
 They do not appear  in static approaches 
 due to their  cancellation against the  wave function renormalization 
 contributions \cite{GaH76}, which have  
 their  origin  in the renormalization of the baryonic states
 when eliminating the mesonic wave function components.
 This concept breaks down  beyond the $\pi$-threshold 
 if full $\pi$-retardation is considered since the  $\pi N N $-component 
can be on-shell.  Therefore, we do not orthonormalize and no
wave function renormalization contributions  appear.
Consequently, the recoil current and   charge densities 
 have to be included.

Besides the $\pi$-MECs and the corresponding Siegert operators, 
our model includes the usual nucleonic one-body current and as most 
important relativistic contribution (besides retardation) the spin-orbit 
current. 
Because  the  $\rho$-mass  is rather large, 
retardation in the $\rho$-MEC is expected to be rather unimportant 
 and therefore not considered in this work.
Concerning the $N\Delta$-$\pi$-MEC, we include retardation in the
corresponding recoil part, 
whereas, due to their minor importance, 
the contact-, meson- and vertex-contributions are
 taken in the static limit.
Since the pion production model of Tanabe and Ohta \cite{TaO85} effectively
incorporates $\omega$-exchange, we include in addition the leading order
 $\rho \pi \gamma$-  and $\omega \pi \gamma$-currents, which are purely 
transverse \cite{RiG97}.

\section{results}
\begin{figure}
\centerline{\psfig{figure=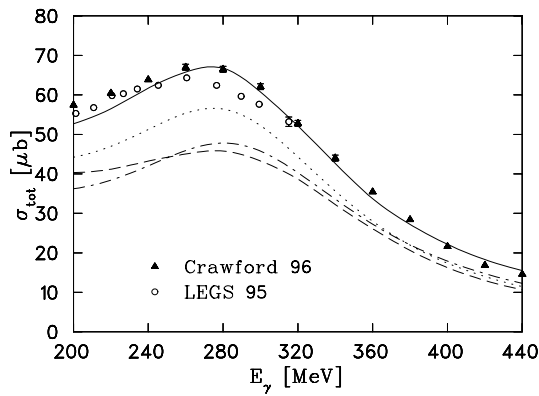,width=7cm,angle=90}}
\vspace{0.5cm}
\caption{Total cross section for $\gamma d \rightarrow p n$ as a function
of photon energy $E_{\gamma}$ in comparison with experiment 
{\protect \cite{{Leg95},{Cra96}}}.
Dashed: static calculation of Wilhelm et al.\ {\protect \cite{Leg95}}:
dotted: static OBEPR-calculation in our approach;
dash-dot: retardation switched on in the hadronic part only,  
 but static MECs;
full: complete calculation including  
 $\pi d$-channel and  $\rho \pi \gamma / \omega \pi \gamma$-MECs.}
\label{sigtot}
\end{figure}

In the numerical evaluation we have included all multipoles and scattering 
 waves up to 
$L = 4$ and $j = 4$. Retardation in the meson-, vertex-, contact-
 and $\rho \pi \gamma / \omega \pi \gamma$-currents 
 is only included up to $L = 2$.
 For the higher multipoles the static limit has been adopted, 
 because inclusion of 
 retardation  effects there does not show significant effects on observables.
 We start the discussion of deuteron photodisintegration with the total
 cross section  shown in Fig.\ \ref{sigtot}. Similar to 
 \cite{{WiA93},{Leg95}}, 
 the static calculation (without $\pi d$-channel and 
 $\rho \pi \gamma  / \omega \pi \gamma$-MECs)
 considerably underestimates the data (dotted curve).
 However, the $\gamma N \Delta$-coupling of \cite{TaO85} used in our 
 approach  is somewhat larger than the one used by Wilhelm et al., 
  so that the 
 discrepancy  between our static calculation  and experiment 
 is not as dramatic as in  \cite{{WiA93},{Leg95}}. 
 The reduction of the cross section by the inclusion of retardation 
 in the hadronic interaction (dash-dotted curve) is, however, 
 overcompensated
 if  retardation in the $\pi$-MECs is added, of which the recoil 
 contributions turn out to  be the most important ones.  
 This destructive interference of retardation in potential and recoil 
 current corresponds to the cancellation of wave function 
 renormalization and recoil current in the static limit. However, here
 remains a net effect leading to a significant increase.
 The cross section is further enhanced by inclusion of  the  $\pi d$-channel
 and  the  $\rho \pi \gamma / \omega \pi \gamma$-MECs
 (by about 8\% respectively 3\%  at 260 MeV) 
 so that the  complete calculation (full curve) agrees quite well
 with  the experimental data over the whole 
 energy range.

 Concerning  the role of the $\pi d$-channel 
 in  $NN$-scattering, its influence 
  on  the inelasticity of the $^1D_2$-channel is similar to \cite{TaO89}.
 However, in contrast to our result of  a slight enhancement,
  the $\pi d$-channel 
 leads in \cite{TaO89} to a reduction of  the cross section in 
 $\gamma d  \rightarrow p n$,  
 the reason for which we are presently investigating.

\begin{figure}
\centerline{\psfig{figure=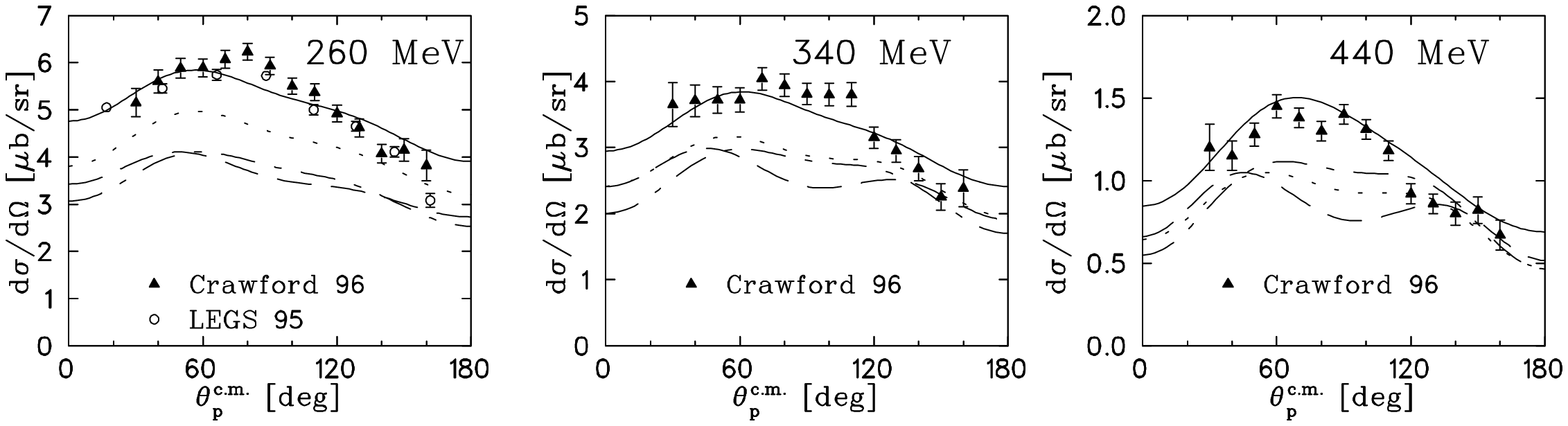,width=17cm,angle=90}}
\vspace{0.5cm}
\caption{Differential cross sections for various energies in comparison with 
         experiment {\protect \cite{{Leg95},{Cra96}}}.
          Notation of the curves as in Fig.\ {\protect \ref{sigtot}}.}  
\label{wqdiff}
\end{figure}

\begin{figure}
\centerline{\psfig{figure=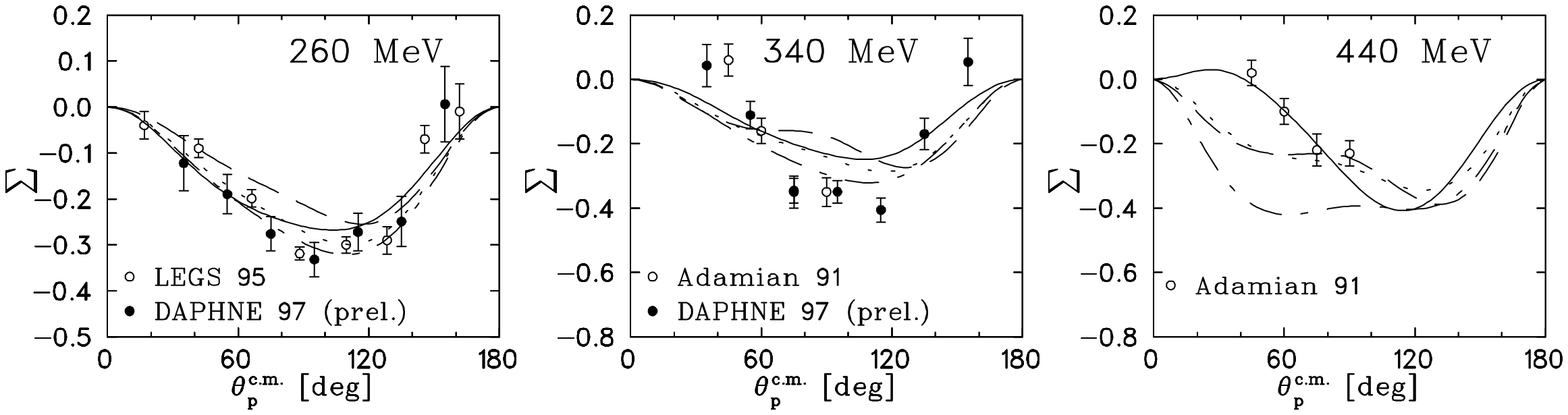,width=17cm,angle=90}}
\vspace{0.5cm}
\caption{Photon asymmetry $\Sigma$ for various energies in comparison with 
         experiment {\protect \cite{{Leg95},{War97},{Ada91}}}.
          Notation of the curves as in 
         Fig.\ {\protect \ref{sigtot}}.}  
\label{sigma}
\end{figure}

 In Figs.\ \ref{wqdiff} and \ref{sigma}, we show differential cross sections
 and photon asymmetries for various energies. Whereas now 
 the differential cross 
 section is in satisfactory agreement with the data, 
 we slightly underestimate
 the absolute size of the asymmetry. In contrast to 
 \cite{{TaO89},{WiA93}} we are able to reproduce quite well 
 the shape of these two  observables also at higher energies,
 in particular the dips in the angular distributions have disappeared.

 In conclusion, we have demonstrated that a CC  approach
 including $N\Delta$- and $\pi d$-channels with full pion 
 retardation in potentials and exchange currents
 is able to remove almost quantitatively  
 the still existing discrepancies between theory
 and experiment in deuteron photodisintegration in the $\Delta$-region.  
 Further improvements of our model should include 
 relativistic contributions to internal dynamics  and currents 
 whereas boost effects 
 are  expected to be small \cite{RiG97}. 
 Furthermore, a realistic $NN$-interaction is desirable  which includes
 isobar degrees of freedom in a coupled channel approach from the 
 beginning so that no box renormalization procedure is necessary.
 The success of our model in $\gamma d \rightarrow NN$
 encourages us to study retardation  in related e.m.\ processes 
 on the deuteron    in the near future, like $\pi$-photoproduction and 
 electrodisintegration.

\centerline{\bf ACKNOWLEDGMENT}
We would like to thank S.\ Wartenberg from the A2 collaboration for providing
 us with the  preliminary results on  the photon asymmetry prior to
 publication \cite{War97}.

\end{document}